\documentclass[aps,prl,showpacs,floatfix,twocolumn,amsmath,amssymb,preprintnumbers]{revtex4-1}
\usepackage{mathrsfs}
\usepackage[figuresright]{rotating}
\usepackage{amsmath}
\usepackage{amssymb}
\usepackage{graphicx}
\usepackage{color}
\usepackage{dcolumn}
\usepackage{bm}
\usepackage[breaklinks=true,colorlinks=true,linkcolor=blue,urlcolor=blue,citecolor=blue]{hyperref}

\usepackage{soul}

\makeatletter
\newcommand{\rmnum}[1]{\romannumeral #1}
\newcommand{\Rmnum}[1]{\expandafter\@slowromancap\romannumeral #1@}
\makeatother

\begin{document}

\title{Realization of Kondo chain in CeCo$_2$Ga$_8$}

\author{Kangqiao Cheng$^{1}$}
\author{Le Wang$^{2,3}$}
\author{Yuanji Xu$^{2,3}$}
\author{Feng Yang$^{1}$}
\author{Haipeng Zhu$^{1}$}
\author{Jiezun Ke$^{1}$}
\author{Xiufang Lu$^{1}$}
\author{Zhengcai Xia$^{1}$}
\author{Junfeng Wang$^{1}$}
\author{Youguo Shi$^{2,3}$}
\email[]{ygshi@iphy.ac.cn}
\author{Yifeng Yang$^{2,3}$}
\email[]{yifeng@iphy.ac.cn}
\author{Yongkang Luo$^{1}$}
\email[]{mpzslyk@gmail.com}
\address{$^1$Wuhan National High Magnetic Field Center and School of Physics, Huazhong University of Science and Technology, Wuhan 430074, China;}
\address{$^2$Beijing National Laboratory for Condensed Matter Physics, Institute of Physics, Chinese Academy of Sciences, Beijing 100190, China;}
\address{$^3$School of Physical Sciences, University of Chinese Academy of Sciences, Beijing 100190, China}

\date{\today}

\begin{abstract}

We revisited the anisotropy of the heavy-fermion material CeCo$_2$Ga$_8$ by measuring the electrical resistivity and magnetic susceptibility along all the principal $\mathbf{a}$-, $\mathbf{b}$- and $\mathbf{c}$-axes. Resistivity along $\mathbf{c}$-axis ($\rho_c$) shows clear Kondo coherence below about 17 K, while both $\rho_{a}$ and $\rho_{b}$ remain incoherent down to 2 K. The magnetic anisotropy is well understood within the theoretical frame of crystalline electric field effect in combination with magnetic exchange interactions. We found the anisotropy ratio of these magnetic exchange interactions, $|J_{ex}^c/J_{ex}^{a,b}|$, reaches a large value of 4-5. We, therefore, firmly demonstrate that CeCo$_2$Ga$_8$ is a quasi-one-dimensional heavy-fermion compound both electrically and magnetically, and thus provide a realistic example of \textit{Kondo chain}.

\end{abstract}


\maketitle

\begin{figure}[!ht]
\vspace*{-10pt}
\hspace*{0pt}
\includegraphics[width=9.5cm]{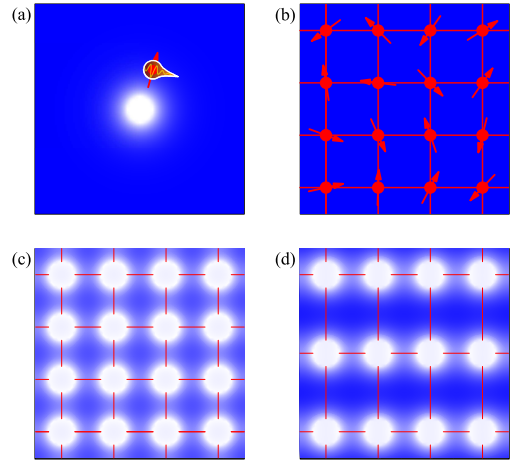}
\vspace*{-5pt}
\caption{\label{Fig1} Schematic diagrams of Kondo effects. (a) A local moment (red) immerses in a metal (blue) forming a Kondo singlet. The spin of the local moment is screened by the conduction electrons and forms a resonance (white haze). (b) A dense lattice of local moments in a metal, without Kondo coupling. (c) Appearance of Kondo coherence for a multi-dimensional dense lattice, leading to heavy-fermion narrow band. (d) A Kondo chain may appear when Kondo coherence is of one dimensional characteristic.}
\end{figure}

A \textit{lattice} in Condensed Matter Physics typically means a periodic repetition and extension in a multi-dimensional space. An example is crystalline lattice, in which the atoms are connected by chemical bondings and construct a stereo network\cite{Ashcroft-SSP}. The concept also applies to more virtual systems, like vortex lattice\cite{Abrikosov-VortexLattice}, skyrmion lattice\cite{Muhlbauer-MnSiSKX2009}, Kondo lattice \textit{et al}. A Kondo singlet, the ``unit cell" of \textit{Kondo lattice}, comes into being when a localized magnetic moment immersed into fermi sea of a metal\cite{Coleman-HFDimension} is quenched by entangling with conduction-electron spin to form a magnetic singlet [Fig.~\ref{Fig1}(a)]. For a dense lattice of local moments [Fig.~\ref{Fig1}(b)], if the Kondo coupling is sufficiently strong, the Kondo singlets communicate to each other and develop a coherent narrow band of which the low-energy quasiparticles are of greatly enhanced effective mass [Fig.~\ref{Fig1}(c)]. The situation may change if the dense lattice reduces to one dimension (1D) or quasi-1D, say, the percolation path of Kondo coherence can be realized only in one direction but fails in other directions [Fig.~\ref{Fig1}(d)]. Systems like this may be termed as a \textit{Kondo chain}.

``Dimensions are critical", as Coleman declared in Ref.~\cite{Coleman-HFDimension}. This is because lower dimension means more phase space for long-wavelength fluctuations and a larger magnetic frustration parameter\cite{Custers-Ce3Pd20Si6QCP}, the latter of which dictates the way that the system undergoes from a quantum ordered state to a disordered state: a conventional spin-density-wave (SDW) type quantum critical point (QCP)\cite{Hertz-QCP,Millis-QCP} or an unconventional Kondo-destruction type QCP\cite{SiQ-localQCP,Coleman-QCP2005,Gegenwart2008}. Kondo destruction generically requires large spin fluctuations and thus favors lower dimension. So far, most known examples of Kondo-destruction QCP were observed in materials between 2D and 3D\cite{Custers-YbRh2Si2QCP,Park-CeRhIn5UnconQCP,Custers-Ce3Pd20Si6QCP,Shishido-CeIn3,Kim-Ce2Pt2Pb,LuoY-CeNiAsOQCP,LuoY-CeNi2As2Pre}, while examples of 1D or quasi-1D have been rare\cite{Krellner-YbNi4P2FMQCP}. The situation in the 1D limit seems elusive: on one hand, long-range magnetic order is hard to stabilize in 1D materials, but on the other hand, a Fermi-surface reconstruction due to $4f$ localized-delocalized transition is still possible according to the predicted global phase diagram for heavy-fermion compounds\cite{Custers-Ce3Pd20Si6QCP}. Extensive material bases are required to elucidate these issues.

Recently, Wang \textit{et al} reported the synthesis and physical properties of CeCo$_2$Ga$_8$\cite{WangL-CeCo2Ga8}. This compound crystalizes in the YbCo$_2$Al$_8$-type orthorhombic structure\cite{Koterlin-Ce128}, in the space group Pbam (No. 55). The crystalline structure of CeCo$_2$Ga$_8$ can be viewed as individual Ce chains along $\mathbf{c}$ axis, and each chain is surrounded by five polyhedral CoGa$_9$ cages in the $\mathbf{ab}$ plane. Since the inter-chain Ce-Ce distances (6.5 and 7.5 \AA) are much longer than the intra-chain distance (4.05 \AA), the compound is considered as a candidate of quasi-1D Kondo lattice\cite{WangL-CeCo2Ga8}. Electrical resistivity, magnetic susceptibility and specific heat measurements manifested that it does not exhibit long-range ordering down to 0.1 K, but show non-Fermi-liquid behavior with a large electronic Sommerfeld coefficient $\gamma_0$$\sim$800 mJ/(mol$\cdot$K$^2$). First-principles calculations revealed flat Fermi sheets arising from itinerant $4f$ band, characteristic of quasi-1D nature. All these suggest that CeCo$_2$Ga$_8$ sits nearby a quantum critical point. However, direct evidence for its quasi-1D nature is still lacking, or equivalently the question is, whether it realizes a Kondo chain at low temperature.

Experimentally, one may expect some physical properties for a Kondo chain compound. Electrically, coherent Kondo scattering would occur only in one direction, or the difference in Kondo-coherence temperature for various directions is huge; and magnetically, before Kondo effect sets in, the magnetic properties and exchange interactions exhibit a great anisotropy. In this paper, we performed careful measurements of resistivity and susceptibility on single crystalline CeCo$_2$Ga$_8$ along all the principal $\mathbf{a}$-, $\mathbf{b}$- and $\mathbf{c}$-axes. We observe all these supposed features for a Kondo chain. Theoretical calculations based on crystalline electric field (CEF) effect were carried out to understand these anisotropies.


Single crystalline CeCo$_2$Ga$_8$ was grown by a Ga self-flux method as described previously\cite{WangL-CeCo2Ga8}. The as-grown samples mostly are needle-like with typical length $\sim$3 mm along $\mathbf{c}$-axis, and about 1$\times$1 mm$^2$ in cross-section. We carefully polished the crystals to make the long-side along $\mathbf{a}$-, $\mathbf{b}$- and $\mathbf{c}$-axes, respectively, and the orientations are verified by X-ray diffraction. These polished single crystals are used for anisotropic electric resistivity and magnetic susceptibility measurements, for which Physical Property Measurement System (PPMS, Quantum Design) and Magnetic Property Measurement System (MPMS, Quantum Design) were employed.


\begin{figure}[t]
\vspace*{-20pt}
\includegraphics[width=8.5cm]{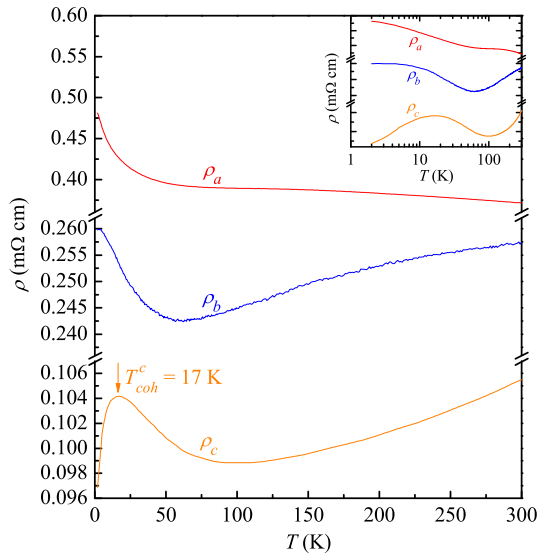}
\vspace*{-15pt}
\caption{\label{Fig2} (a) Temperature dependence of electrical resistivity of CeCo$_2$Ga$_8$ for current applied along $\mathbf{a}$-, $\mathbf{b}$- and $\mathbf{c}$-axes, respectively. Coherent Kondo scattering behavior is only seen in $\rho_c$ for $T$ below 17 K. Inset shows $\rho(T)$ in a semi-logarithmic frame.}
\end{figure}

We start with electrical resistivity, as shown in Fig.~\ref{Fig2}. Here the resistivities were measured for electrical currents along all $\mathbf{a}$-, $\mathbf{b}$- and $\mathbf{c}$-axes, and results are denoted as $\rho_a$, $\rho_b$ and $\rho_{c}$, respectively. At room temperature, the resistivity values respectively are 0.372, 0.257 and 0.106 m$\Omega\cdot$cm. Upon cooling, both $\rho_b$ and $\rho_c$ decrease, while $\rho_a$ slightly increases. A tiny bump is visible in both $\rho_a(T)$ and $\rho_b(T)$ for temperatures near 170 K, which is probably due to spin-flip scattering by the excitations between CEF levels. We will discuss the CEF effect later on in detail. A common feature for lower temperature is that below $\sim$50 K all $\rho(T)$ curves turns up logarithmically, reminiscent of incoherent Kondo scattering\cite{Kondo-RMinimum}. In the low temperature limit, $\rho_c(T)$ turns down, manifesting the establishment of Kondo coherence. The coherence temperature can be defined as $T^c_{coh}$$\approx$17 K. Our previous work has systematically studied $\rho_c$ under high pressure\cite{WangL-CeCo2Ga8}. The sub-Kelvin measurements  there revealed non-Fermi-liquid behavior ($\rho_c$$\propto$$T$) at low temperature, and a recovery of Fermi liquid ($\rho_c$$\propto$$T^2$) under pressure, implying that the compound sits in the vicinity of a quantum critical regime. In contrast, both $\rho_a$ and $\rho_b$ remain incoherent down to 2 K. It should be noted that $\rho_b(T)$ tends to level off below 3 K. It is possible that Kondo coherence may occur below 2 K. Any way, $\rho_a$ and $\rho_b$ require further investigations in the future. Noteworthy that the previously known quasi-1D Kondo-lattice candidate YbNi$_4$P$_2$ actually displays coherent Kondo scatterings for currents along both inter-chain and intra-chain\cite{Krellner-YbNi4P2Rho}. This places CeCo$_2$Ga$_8$ in a regime with even lower dimension than YbNi$_4$P$_2$.

\begin{figure}[t]
\vspace*{-20 pt}
\includegraphics[width=8.5cm]{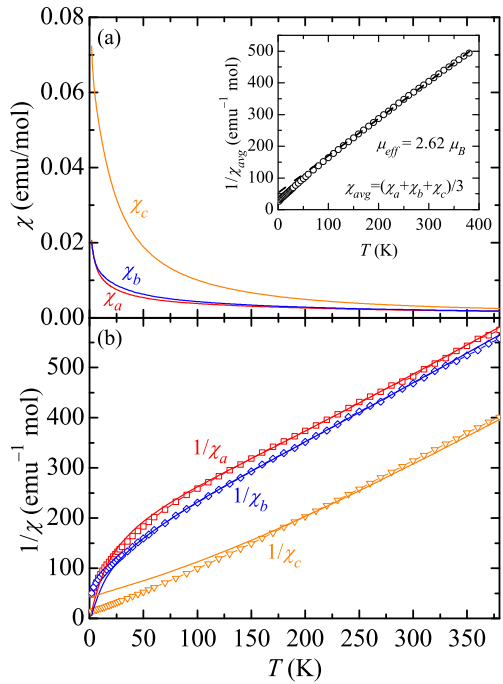}
\vspace*{-20pt}
\caption{\label{Fig3} (a) Temperature dependent magnetic susceptibilities of CeCo$_2$Ga$_8$ measured for external field $\mathbf{B}$ parallel to $\mathbf{a}$-, $\mathbf{b}$- and $\mathbf{c}$-axes, respectively. All the measurements were made under $B$=0.1 T in zero-field cooling protocol. The inset shows Curie-Weiss behavior of $\chi_{avg}(T)$, where $\chi_{avg}$=($\chi_a$$+$$\chi_{b}$$+$$\chi_c$)/3 is the powder-average susceptibility. (b) Inverse susceptibilities as a function of temperature. The symbols stand for experimental data, while the solid lines are theoretically calculated results based on CEF theory.}
\end{figure}

Turning now to magnetic susceptibilities ($\chi$) as shown in Fig.~\ref{Fig3}. The magnetic anisotropy for field parallel and perpendicular to $\mathbf{c}$ has been reported in our earlier paper\cite{WangL-CeCo2Ga8}, while the in-plane anisotropy remains unknown. This is now complemented in Fig.~\ref{Fig3}. We find generically $\chi_c$$>$$\chi_b$$\gtrsim$$\chi_a$ over the full temperature range, only except for below 5 K where $\chi_a(T)$ and $\chi_b(T)$ slightly cross. The anisotropic ratio $\chi_c/\chi_{a,b}$ reaches 3.5 at 2 K. We calculated the powder-average susceptibility $\chi_{avg}$=($\chi_a$$+$$\chi_{b}$$+$$\chi_c$)/3, the inverse of which has been displayed in the inset to Fig.~\ref{Fig3}(a). By fitting $\chi_{avg}(T)$ to a standard Curie-Weiss law, we derive the effective moment $\mu_{eff}$$=$2.62 $\mu_B$/f.u. This value is relatively larger than that of a free Ce$^{3+}$ (2.54 $\mu_B$), suggesting that the Co ions also carry some magnetic moment. It is very common that Co ions in Ce-contained compounds are magnetic, see for e.g. CeCoAsO\cite{Sarkar-CeCoAsO} and CeCo$_2$As$_2$\cite{Thompson-RCo2As2}.

It is well known that the CEF effect plays a key role in the anisotropic magnetism of rare earth ions. The CEF Hamiltonian for a $D_{2h}$ point group contains five terms\cite{Bauer-CEF},
\begin{equation}
\hat{H}_{CEF}=B_2^0 \hat{O}_2^0+B_2^2 \hat{O}_2^2+B_4^0 \hat{O}_4^0+B_4^2 \hat{O}_4^2+B_4^4 \hat{O}_4^4,
\label{Eq1}
\end{equation}
where $B_l^m$ ($l$=2,4; $m$=0,2,4) are the CEF parameters, and $\hat{O}_l^m$ are Steven¡¯s operators\cite{Stevens-Operators,Hutchings-CEF}. Note that $\hat{O}_2^2$ and $\hat{O}_4^2$ turn on the in-plane anisotropy. In addition, the Zeeman effect and exchange interaction should also be taken into account,
\begin{eqnarray}
\hat{H}_{Zee} = -g \mu_B \hat{\textbf{J}}\cdot\textbf{B},~~~~~~~~~~~~~~~~~~\label{Eq2}\\
\hat{H}_{ex}=-\sum_{<i,j>}(J_{ex}^{a}S_i^xS_j^x+J_{ex}^{b}S_i^yS_j^y+J_{ex}^{c}S_i^zS_j^z),\label{Eq3}
\end{eqnarray}
where $\hat{\mathbf{J}}$=($\hat{J}_x$,$\hat{J}_y$,$\hat{J}_z$) is the total angular momentum operator, $g$=6/7 is the Land\'{e} factor of Ce$^{3+}$ ions, $\mu_B$ is Bohr magneton, $J_{ex}^{a,b,c}$ are the components of the nearest-neighbor exchange interaction with Ce$^{3+}$ moment along $\mathbf{a}$, $\mathbf{b}$ and $\mathbf{c}$, respectively.

An exact calculation of the paramagnetic susceptibility in the presence of CEF and magnetic interactions have been performed by P. Boutron\cite{Boutron-PMchi}. At high temperature, a series expansion of $\chi$ in $1/T$ leads to the expressions of $1/\chi(T)$ as following,
\begin{eqnarray}
\begin{aligned}
\frac{1}{\chi_{\alpha}}&=\frac{1}{C}[T-(\theta_{CEF}^{\alpha}+\theta_{ex}^{\alpha})]\\
&=\frac{1}{\chi_{\alpha}^{CEF}}-\frac{\theta_{ex}^{\alpha}}{C},~~(\alpha=a,b,c)\label{Eq4}
\end{aligned}
\end{eqnarray}
in which $C$ is the Curie constant, and $\theta_{ex}^{\alpha}$=$\frac{j(j+1)}{3}J_{ex}^{\alpha}$ with $j$=5/2 for Ce$^{3+}$. This approximation turns out to be rather good for the temperature region $T$$\gg$$|\theta_{ex}^{\alpha}|$\cite{Cho-HoNi2B2C_CEF,Gingras-Tb2Ti2O7_CEF,LuoY-CeNi2As2}. Eq.~(\ref{Eq4}) tells us that the total Weiss temperature is a sum of both $\theta_{CEF}$ and $\theta_{ex}$. The way to get the intrinsic magnetic correlation is to separate out $\theta_{ex}$, as $\theta_{CEF}$ usually is more dominant. We numerically calculated $\chi^{CEF}$, and fitted to our experimental data. The results are shown in $1/\chi$ in Fig.~\ref{Fig3}(b). The best fitting parameters include $B_l^m$ that have been summarized in Table \ref{Tab1}. The calculation also results in the exchange interactions $J_{ex}^a$$\approx$2.5 K, $J_{ex}^b$$\approx$3 K, $J_{ex}^c$$\approx$$-$12 K. Note that the signs of $J_{ex}^c$ and $J_{ex}^{a,b}$ are opposite. This indicates that the magnetic correlations are probably antiferromagnetic for intra-chain, but are ferromagnetic for inter-chain. The anisotropy ratio of magnetic interactions can thus be estimated $|J_{ex}^{c}/J_{ex}^{a,b}|$$\sim$4-5. It, therefore, is reasonable to draw a conclusion that CeCo$_2$Ga$_8$ is magnetically quasi-1D. It should be emphasized that the strongest exchange interaction, $J_{ex}^c$, is comparable to but relatively smaller than $T_{coh}^{c}$, which is consistent with the fact that the compound resides on the edge of a magnetic instability\cite{WangL-CeCo2Ga8}.

We should admit that at low temperature, the fittings are not excellent. There are several reasons for such deviation. (\rmnum{1}) At low temperatures, the condition of $T$$\gg$$\theta_{ex}^{\alpha}$ is not satisfied, especially for $\mathbf{c}$-axis, therefore the approximation of Eq.~(\ref{Eq4}) is no longer valid. (\rmnum{2}) Kondo effect becomes more and more important at low temperature. This is particularly the case for $\mathbf{c}$-axis, as Kondo coherence is seen in $\rho_c$ below 17 K. (\rmnum{3}) As the 1D spin-chain nature becomes more prominent, $\chi(T)$ no longer simply obeys the Curie-Weiss law\cite{Bonner-1DChian,WangL-CeCo2Ga8}, which is not considered in the CEF theory, either. (\rmnum{4}) The CEF parameters $B_l^m$ can be temperature dependent, but is beyond the scope of the present work. Despite of these shortcomings, the agreement between calculations and experiments is decent.

\begin{table*}
\caption{\label{table2} CEF parameters, energy levels and wave functions in CeCo$_2$Ga$_8$ at zero magnetic field.}
\label{Tab1}
\begin{ruledtabular}
\begin{center}
\def\temptablewidth{1.6\columnwidth}
\begin{tabular}{ccccccc}
\multicolumn{6}{l}{CEF parameters } \\
 & $B_2^0$=$-$22.0(5) K, & $B_2^2$=0.8(1) K, & $B_4^0$=0.6(1) K, & $B_4^2$=0.45(5) K, & $B_4^4$=1.4(1) K \\
 \\ \hline \hline
\multicolumn{6}{l}{Energy levels and Eigenstates}  \\  $E$(K)& $\mid
\frac{5}{2},+\frac{5}{2} \rangle$ & $\mid \frac{5}{2},+\frac{3}{2}
\rangle$ & $\mid \frac{5}{2},+\frac{1}{2} \rangle$ & $\mid
\frac{5}{2},-\frac{1}{2} \rangle$
& $\mid \frac{5}{2},-\frac{3}{2} \rangle$ & $\mid \frac{5}{2},-\frac{5}{2} \rangle$ \\
\hline
0       &       0     &  $-$0.2763  &      0      &  $-$0.0371  &      0      &     0.9604  \\
0       &     0.9604  &      0      &  $-$0.0371  &     0       &  $-$0.2763  &     0       \\
145(5)  &       0     &     0.9609  &      0      &     0.0055  &      0      &     0.2767  \\
145(5)  &     0.2767  &      0      &     0.0055  &     0       &     0.9609  &     0  \\
445(5)  &     0.0341  &      0      &     0.9993  &     0       &  $-$0.0156  &     0  \\
445(5)  &     0       &  $-$0.0156  &     0       &     0.9993  &      0      &     0.0341  \\
\end{tabular}
\end{center}
\end{ruledtabular}
\end{table*}

\begin{figure}[t]
\vspace*{-5pt}
\hspace*{-10pt}
\includegraphics[width=9.0cm]{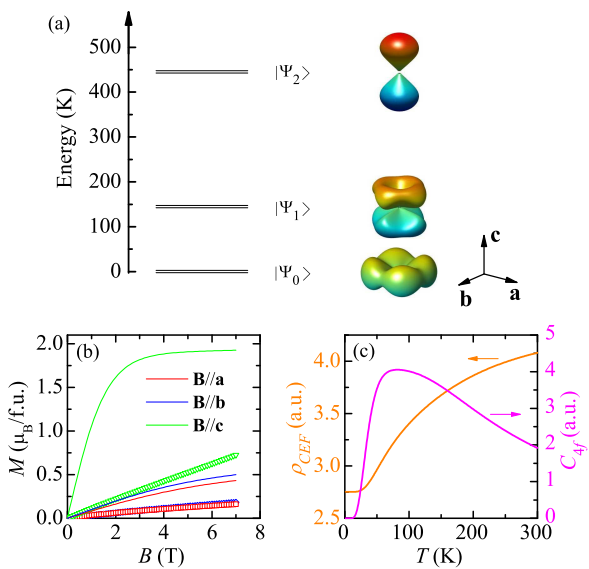}
\vspace*{-25pt}
\caption{\label{Fig4} (a) Sketch of CEF splitting of CeCo$_2$Ga$_8$, as well as the contours of $4f$ charge density for each doublet. The calculations were based on the parameters listed in Table \ref{Tab1}. (b) Isothermal field dependence of magnetization at 2 K. The symbols denote experimental data, and the solid lines signify simulated ones based on CEF theory \textit{without} magnetic exchange interactions. The colours are: red-$\mathbf{a}$, blue-$\mathbf{b}$, and green-$\mathbf{c}$. (c) Calculated CEF resistivity (left) and specific heat (right) as functions of temperature.}
\end{figure}

A schematic of the CEF splitting is presented in Fig.~\ref{Fig4}(a). The $j$=5/2 multiplet splits into three Kramers doublets, with the first and second excited doublets sitting at $\sim$145 K and 445 K above the ground states. The ground state can be expressed in term of $|\Psi_0\rangle$=$\alpha |\pm5/2\rangle$$+$$\sqrt{1-\alpha^2-\beta^2}|\mp3/2\rangle$$+$$\beta |\pm1/2\rangle$, with $\alpha$=0.9604 and $\beta$=$-$0.2763. The large $\alpha^2$($\gg$1/6) makes the orbital severely oblate\cite{Willers-Ce115CEF}, see the calculated contour of the $4f$ charge distribution in Fig.~\ref{Fig4}(a) \cite{Freeman-RareEarth,Bauer-CEF}. For Ce$^{3+}$, an oblate orbital typically renders a magnetic easy-axis. This can be seen from the simulated isothermal magnetizations [Fig.~\ref{Fig4}(b)]. Since in this simulation magnetic exchange interactions and Kondo effect are not considered, a direct comparison with experimental results is not perfect, but $\mathbf{c}$ axis being the magnetic easy-axis is inescapable. Intuitively, this should be an important premise to form a quasi-1D magnetic system.

Going a little further, we also calculated the CEF contribution to electrical resistivity ($\rho_{CEF}$) and specific heat ($C_{4f}$), the results of which are displayed in Fig.~\ref{Fig4}(c). $\rho_{CEF}$, as we mentioned before, arising from spin-flip scattering by CEF excitations, shows an obvious hump near 150 K, which is in agreement with the experimental $\rho_a$ and $\rho_b$ (Fig.~\ref{Fig2}) (Note \cite{note1}). This provides additional evidence for the validity of our analysis. A broad peak near 75 K is expected in $C_{4f}(T)$, \textit{i.e.} Schottky anomaly. A direct comparison to experiment is not available now, because all our attempts to synthesize the non-magnetic counter-part LaCo$_2$Ga$_8$ failed, and we are not able to extract $C_{4f}(T)$ from the total specific heat of CeCo$_2$Ga$_8$.


To draw a conclusion, we carefully measured the electrical resistivity and magnetic susceptibility of the heavy-fermion compound CeCo$_2$Ga$_8$ along all three principal $\mathbf{a}$-, $\mathbf{b}$- and $\mathbf{c}$-axes. Anisotropies can be seen in both resistivity and susceptibility. While $\rho_c$ displays coherent Kondo scattering below 17 K, the Kondo scattering in $\mathbf{a}$ and $\mathbf{b}$ remain incoherent down to 2 K. The magnetic anisotropy can be well understood by theoretical calculations based on CEF theory. The calculations confirm $\mathbf{c}$-axis as the easy axis, and also derives an anisotropy ratio of magnetic exchange interactions $|J_{ex}^c/J_{ex}^{a,b}|$$\sim$4-5. We, therefore, firmly demonstrate that CeCo$_2$Ga$_8$ is a quasi-one-dimensional heavy-fermion compound both electrically and magnetically, and thus provide a realistic example of Kondo chain. This finding has an immediate ramification for the global phase diagram of heavy-fermion materials\cite{Custers-Ce3Pd20Si6QCP}. Our previous specific heat measurements suggested unconventional quantum critical behavior in CeCo$_2$Ga$_8$\cite{WangL-CeCo2Ga8}. Altogether, a Kondo-destruction QCP seems possible in the quasi-1D limit. This, of course, requires more evidence to clarify. One should also be noted that the YbCo$_2$Al$_8$ family has many other Ce128 counterparts\cite{Koterlin-Ce128} whose physical properties remain unclear. They open an important means to explore heavy-fermion materials for studies on quantum criticality and dimensionality.

Finally, it is worthwhile to mention that, although the system on the whole behaves quasi-1D electrical and magnetic properties, the inter-chain coupling along $\mathbf{b}$ seems relatively stronger than along $\mathbf{a}$. This can be seen from many aspects of view. The magnitude of $\rho_b$ is smaller than that of $\rho_{a}$, and moreover, coherence-like Kondo scattering seems to be on the verge in the former. Magnetically, $\chi_b$ is a little larger than $\chi_a$, and most importantly, the estimated exchange interaction is also slightly stronger in $\mathbf{b}$. A weak in-plane anisotropy is likely.


We thank Hua Chen for helpful discussion. Y. Luo is supported by the 1000 Youth Talents Plan of China. The authors acknowledge National Natural Science Foundation of China (Grant No. 11674115, 11522435, 11774399 and 11474330) and the National Key R\&D Program of China (Grant No. 2017YFA0303103, 2016YFA0401704, 2017YFA0302901 and 2016YFA0300604).


\begin{thebibliography}{34}
\bibitem [{not()}]{note1}%
  \BibitemOpen
  \href@noop {} {}\bibinfo {note}
  {$\rho_{CEF}$$\propto$$\sum\limits_{m_s,m_s'}\sum\limits_{i,i'} (\langle
  m_s';\Psi_{i'}|(\hat{s}_z\hat{J}_z$$+$$\hat{s}_+\hat{J}_-$$+$$\hat{s}_-\hat{J}_+)|m_s;\Psi_{i}
  \rangle)^2$, where $\hat{\mathbf{s}}$ is the spin operator of conduction
  electrons, and $m_s$=$\pm$1/2. Since $\mathbf{c}$-axis is the magnetic easy
  axis for CeCo$_2$Ga$_8$, $\rho_{c}$ is expected to be little affected by CEF.
  This is consistent with our experiments}\BibitemShut {NoStop}%
\end{thebibliography}
%

\end{document}